\documentclass{article}
\usepackage{waspaa19_aut,amsmath,url,times}

\usepackage{color}
\usepackage[pdftex]{graphicx}

\usepackage{xcolor}
\usepackage[font=small,skip=9pt]{caption}
\usepackage{balance}

\newcommand\myeq{\mkern1.5mu{=}\mkern1.5mu}

\usepackage[pscoord]{eso-pic}
\newcommand{\placetextbox}[3]{
  \setbox0=\hbox{#3}
  \AddToShipoutPictureFG*{
    \put(\LenToUnit{#1\paperwidth},\LenToUnit{#2\paperheight}){\vtop{{\null}\makebox[0pt][c]{#3}}}%
  }%
}%

\title{An Improved Measure of Musical Noise Based on Spectral Kurtosis}

\name{Matteo Torcoli}
\address{Fraunhofer Institute for Integrated Circuits IIS, Erlangen, Germany\\matteo.torcoli@iis.fraunhofer.de\\                
}

\begin{document}

\ninept
\maketitle

\begin{sloppy}

\begin{abstract}
Audio processing methods operating on a time-frequency representation of the signal can introduce unpleasant sounding artifacts known as musical noise. These artifacts are observed in the context of audio coding, speech enhancement, and source separation. The change in kurtosis of the power spectrum introduced during the processing was shown to correlate with the human perception of musical noise in the context of speech enhancement, leading to the proposal of measures based on it. These baseline measures are here shown to correlate with human perception only in a limited manner. 
As ground truth for the human perception, the results from two listening tests are considered: one involving audio coding and one involving source separation. 
Simple but effective perceptually motivated improvements are proposed and the resulting new measure is shown to clearly outperform the baselines in terms of correlation with the results of both listening tests.
Moreover, with respect to the listening test on musical noise in audio coding, the exhibited correlation is nearly as good as the one exhibited by the Artifact-related Perceptual Score (APS), which was found to be the best objective measure for this task. The APS is however computationally very expensive. The proposed measure is easily computed, requiring only a fraction of the computational cost of the APS.
\end{abstract}


\placetextbox{0.5}{0.08}{\fbox{\parbox{\dimexpr\textwidth-2\fboxsep-2\fboxrule\relax}{\footnotesize \centering © 2019 IEEE.  Personal use of this material is permitted.  Permission from IEEE must be obtained for all other uses, in any current or future media, including reprinting/republishing this material for advertising or promotional purposes, creating new collective works, for resale or redistribution to servers or lists, or reuse of any copyrighted component of this work in other works. Digital Object Identifier (DOI): 10.1109/WASPAA.2019.8937195}}}%

\section{Introduction}
Musical noise (sometimes also referred to as \textit{musical artifacts}, \textit{musical tones}, or \textit{birdies}) is an audio distortion caused by isolated holes and/or peaks in the power spectrum of the signal. 
These can be introduced by audio processing methods operating on a time-frequency representation of the signal, e.g., audio coding, speech enhancement, or source separation. 
Musical noise is an annoying distortion for the listener, who perceives it as unnatural and warbling fluctuations~\cite{daudet:2003}.
Countermeasures and improved methods were developed (e.g., \cite{ephraim:1984,esch:2009}) as well as objective measures trying to predict the perceived audio quality degradation, e.g., \cite{Hamon:2017}.

In \cite{torcoli:2018comparing}, the Artifact-related Perceptual Score (APS) from the Perceptual Evaluation methods for Audio Source
Separation (PEASS) toolbox \cite{emiya:2011} was shown to correlate well with the perception of musical noise (Pearson's linear correlation  $r=0.95$). 
However, the calculation of the APS is computationally very expensive. This can be problematic, e.g., if the measure is used in an iterative approach which adjusts the parameters of an algorithm so to maximize the quality of each output signal. 

Measures based on the spectral kurtosis (Sec.\,\ref{sec:LKR}) constitute a computationally lighter alternative to APS. Spectral kurtosis and human perception of musical noise were shown to be correlated in~\cite{uemura:2008, yu:2011}. However, in \cite{torcoli:2018comparing} only a limited correlation was found for a previously proposed kurtosis-based measure. 

A more in-depth correlation analysis is carried out in this paper (Sec.\,\ref{sec:evaluation}). 
Simple but effective perceptually motivated improvements are proposed (Sec.\,\ref{sec:newProposal}). The resulting measure is shown to outperform the previously proposed kurtosis-based measures in terms of correlation with perceptual scores. In some cases, the new measure performs as well as the APS at a fraction of the computational cost.
 
 \section{State of the Art}  
\label{sec:LKR}

\subsection{Kurtosis: Definition and General Interpretation}
\label{subsec:def}

In probability theory, the kurtosis $\Psi_X$ of a random variable $X$ is defined as the fourth standardized moment:
\begin{equation}
\Psi_X = \frac{\mu_{4}}{\mu_{2}^{2}} = \frac{\mathrm{E}\left\{ \left(X-\mu\right)^{4}\right\} }{\left(\mathrm{E}\left\{ \left(X-\mu\right)^{2}\right\} \right)^{2}},
\end{equation}
where $\mathrm{E}\left\{ \cdot  \right\}$ is the expectation operator, $\mu_n$ is the $n$-th order central moment of $X$ (e.g., $\mu_2$ is the variance of $X$), and $\mu$ is the mean of $X$, i.e., $\mu=\mathrm{E}\left\{X\right\}$. 

$\Psi_X$ is a variance-normalized measure of the weight of the tail(s) of the distribution of $X$, i.e., a measure of the tendency towards generating outliers.
$\Psi_X$ does not say anything about the shape or the weight of the peak of the distribution of $X$~\cite{westfall:2014}. 

\subsection{Spectral Kurtosis and Musical Noise}
\label{subsec:original_LKR}

In the context of noise reduction for speech enhancement, it was observed that the perception of the introduced musical noise is correlated with the change in kurtosis between the input and the output noise power spectra \cite{uemura:2008}. It is important to note that the kurtosis \textit{change} is considered and not the absolute value of the kurtosis itself. 
Also, only the noise signals are considered. An ideal speech enhancement algorithm would increase the spectral kurtosis while speech is active and leave it unchanged while only noise is present.
Hence, it was proposed to quantify the change in spectral kurtosis in the noise signals by means of the log-kurtosis ratio $\Delta\Psi$ \cite{uemura:2008}:
\begin{equation}
\Delta\Psi = \ln \left(  \frac{ \Psi_{Nout} }{ \Psi_{Nin} }  \right),
\end{equation}
where $Nin$ and $Nout$ are the power spectra of the noise signal before and after processing respectively. 

In \cite{miyazaki:2012}, conditions based on $\Delta\Psi$ were formulated for a musical-noise-free processing and used to derive the optimal parameters for an iterative spectral subtraction algorithm.

In~\cite{uemura:2008} and \cite{miyazaki:2012} as well as in other works from the same group of authors (e.g., \cite{uemura:2009, inoue:2011, saruwatari:2013}), speech and noise power spectra are assumed to be gamma-distributed. Complete knowledge of the internal variables of the processing is also assumed and $\Delta\Psi$ is analytically derived for each case.

\subsection{Black-Box Approach}
\label{subsec:blackbox}

In \cite{yu:2011} it is proposed to use the statistical counterpart to $\Psi_X$, i.e., to replace the population moments with the sample moments so to obtain an instantaneous kurtosis $kurt_X(\ell)$:
\begin{equation}
kurt_X(\ell)=\frac{ \frac{1}{K} \sum_{k=1}^{K} \left(X(k,\ell) -  \overline{X(k,\ell)} \right)^{4} }{\left( \frac{1}{K} \sum_{k=1}^{K} \left(X(k,\ell) - \overline{X(k,\ell)}  \right)^{2} \right)^{2}},
\label{eq:inst_kurt}
\end{equation}
where $X$ is a generic power spectrum, $X(k,\ell)$ is its $k$-th frequency bin at the $\ell$-th time frame, $K$ is the number of bins, and $\overline{X(k,\ell)}$ is the instantaneous sample mean of $X$, i.e., $\overline{X(k,\ell)}=\frac{1}{K}\sum_{k=1}^{K} X(k,\ell)$. 
Hence, $kurt_{Nin}(\ell)$ and $kurt_{Nout}(\ell)$ are obtained by replacing $X$ with $Nin$ and $Nout$ in Eq.\,\ref{eq:inst_kurt}.
The average over all frames $\ell$ is thus computed, obtaining $kurt_{Nin}$ and $kurt_{Nout}$. The black-box log-kurtosis ratio $\Delta kurt$ can finally be computed \cite{yu:2011}:
\begin{equation}
\Delta kurt = \ln \left(  \frac{ kurt_{Nout} }{ kurt_{Nin} }  \right).
\label{eq:deltakurt_Yu}
\end{equation}
This measure is considered a \textit{black box} because it simply needs the signal before and after processing. No assumptions on the distribution of the signal power spectra are made, 
contrary to \cite{uemura:2008, miyazaki:2012, uemura:2009, inoue:2011, saruwatari:2013}.
Moreover, in \cite{yu2012itg} it is proposed to replace Eq.\,\ref{eq:inst_kurt} by a weighted instantaneous spectral kurtosis $kurt^W_X(\ell)$:
\begin{equation}
kurt^W_X(\ell)=\frac{ \frac{1}{K} \sum_{k=1}^{K} \left(X_\alpha(k,\ell) -  \overline{X_\alpha(k,\ell)} \right)^{4} }{\left( \frac{1}{K} \sum_{k=1}^{K} \left(X_\alpha(k,\ell) - \overline{X_\alpha(k,\ell)}  \right)^{2} \right)^{2}},
\end{equation}
where $X_\alpha(k,\ell) = \alpha_X(k) X(k,\ell)$, and $\alpha_X(k)$ represents the signal-and-frequency-dependent weights computed as \mbox{$\alpha_X(k)=\left(  \frac{1}{L} \sum_{\ell=1}^{L} X(k,\ell) \right)^{-1}$}, where $L$ is the total number of time frames.
The resulting weighted log-kurtosis ratio $\Delta kurt^W$ (obtained by replacing $kurt_X$ with $kurt^W_X$ in Eq.\,\ref{eq:deltakurt_Yu}) is reported to correlate with perceptual scores with Pearson's linear correlation $r=0.95$ \cite{yu2012itg}. Yet this is evaluated using only 12 scores spanning the entire quality range, i.e., from \textit{intolerably audible musical noise} to \textit{inaudible musical noise}. A more extensive correlation analysis is carried out in Sec.\,\ref{sec:evaluation}.
The $\Delta kurt^W$ measure is also recommended in~\cite{itut1130:2015}.

\section{The Proposed New Measure}  
\label{sec:newProposal}

This section proposes a new kurtosis-based measure of musical noise. Following a perceptually motivated pre-processing, modified equations for the calculation of the log-kurtosis ratio are proposed. The pre-processing comprises conversion to A-weighted decibels (dBA), limitation of the lower values, and multi-band analysis.

Fig.\,\ref{fig:explain} supports the explanation by visualizing an example consisting of one audio excerpt of a harp arpeggio. 
The analysis takes place in the short-time Fourier-transform domain using a sine window of 1024 samples, 50\% overlap at sampling rate $48$ kHz, and discrete Fourier transform length of double the window length.

\begin{figure}[t]
\centering
\centerline{\includegraphics[width=0.87\columnwidth]{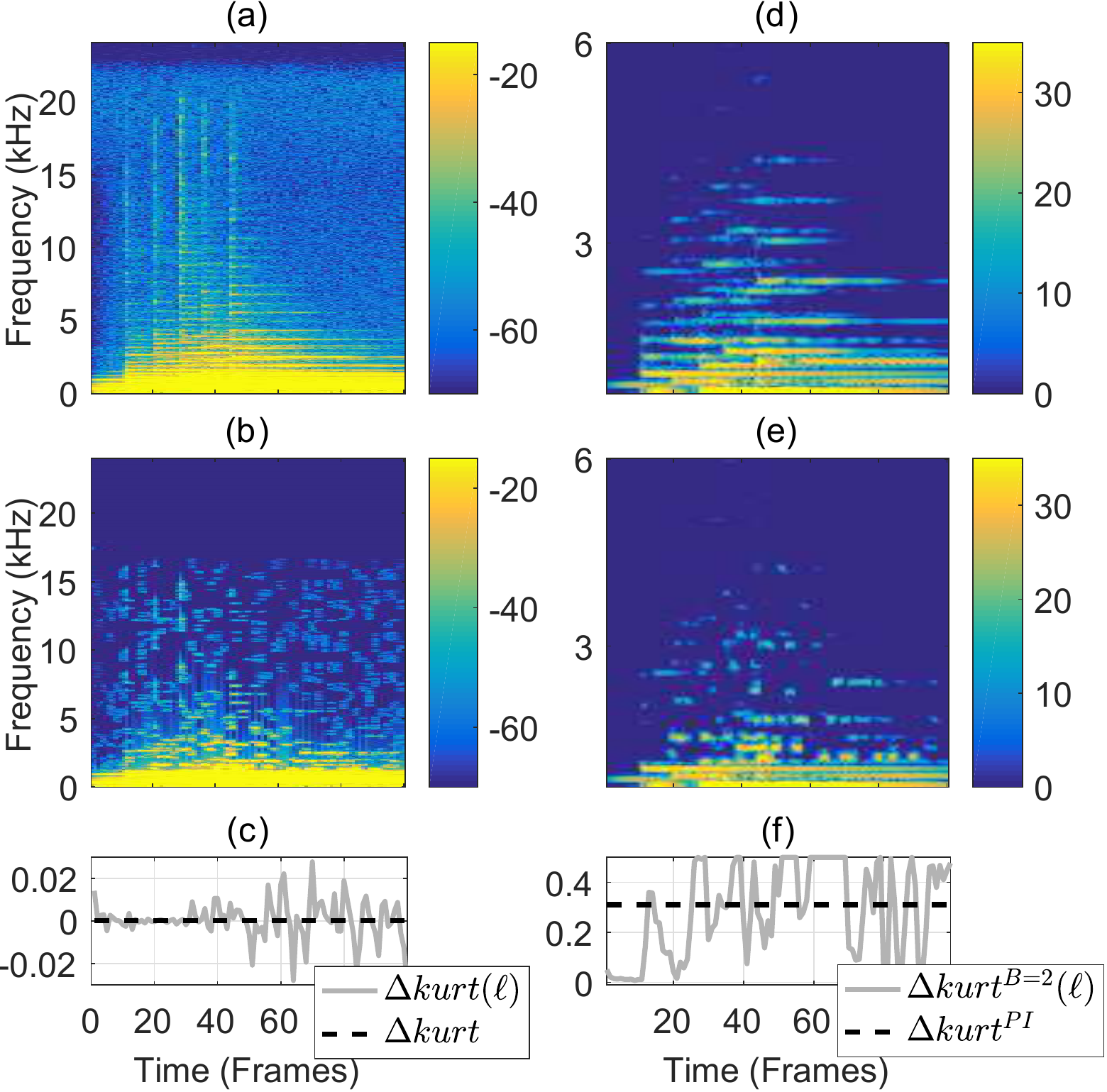}}
\caption{\label{fig:explain}Spectrogram of an arpeggio played on a harp, before ($Nin$, subplot \textbf{(a)}) and after ($Nout$, subplot \textbf{(b)}) introducing musical noise.
Subplot \textbf{(c)} depicts $\Delta kurt(\ell)$ and $\Delta kurt$. 
Subplots \textbf{(d)} and \textbf{(e)} show the sub-band $B\myeq 2$ for $Nin$ and $Nout$, respectively, after the proposed perceptually motivated pre-processing, which comprises conversion to dBA and limitation of the lower values. The resulting $\Delta kurt^{B=2}(\ell)$  and $\Delta kurt^{PI}$ are depicted in subplot \textbf{(f)}. The sub-band $B\myeq 2$ corresponds to $(0.75, 6]$ kHz and it is automatically selected by the algorithm, as it maximizes $\sum_{\ell=1}^{L} \left(  w^B(\ell) \Delta kurt^{B} (\ell) \right)$ for this example.
}
\end{figure}

The spectrogram of the unprocessed signal $Nin$ is depicted in Fig.\,\ref{fig:explain}(a). The processing under test randomly quantizes to zero a fixed percentage of frequency bins. The output spectrogram $Nout$ is depicted in Fig.\,\ref{fig:explain}(b). The introduced holes (and peaks) result in clearly perceivable musical noise. However, Fig.\,\ref{fig:explain}(c) shows that $\Delta kurt$ (Eq.\,\ref{eq:deltakurt_Yu}) is zero, i.e., the previously proposed $\Delta kurt$ fails to detect the musical noise. This happens because the calculation of the spectral kurtosis is dominated by the large number of bins with zero or negligible magnitude. In fact, every bin is identically treated by the previously proposed measures and no aspect of the human auditory system is considered.
The proposed pre-processing (Sec.\,\ref{subsec:prepro}) targets this problem by focusing on perceptually relevant portions of the spectrogram where differences are evident.

\subsection{Perceptually Motivated Pre-Processing}
\label{subsec:prepro}

It is first proposed to convert the spectral power values to dBA, i.e., the generic power spectrum $X(k,\ell)$ is converted to $X_{dBA}(k,\ell)$. This has a dual effect. 
The logarithmic compression better represents the dynamic range of the human ear, while the A-weighting approximates our frequency-dependent sensitivity to sounds.
Then, $X_{dBA}(k,\ell)$ is limited and shifted so to be non-negative ($^{+}$):
\begin{equation}
X^{+}(k,\ell) = \max(X_{dBA}(k,\ell), thr) - thr
\end{equation}
where $thr$ is a threshold on the dBA power values. The values below $thr$ are supposed to correspond to silence or negligible levels. In this implementation $thr = P_{dBA} - 20$, where $P_{dBA}$ is the overall root mean square (RMS) power of $X_{dBA}(k,\ell)$. 
The frames $\ell$ for which $N_{out}^{+}(k, \ell)=0,$ $\forall k$ are discarded.

Hence, the so-obtained $N_{in}^{+}(k, \ell)$ and $N_{out}^{+}(k, \ell)$ are analyzed into sub-bands. In this implementation, 3 sub-bands are used, namely a lower band $(50, 750]$ Hz, a central 3-octaves band $(0.75, 6]$ kHz, and a higher band $(6, 16]$~kHz. 
Frequency outside the range \mbox{$(50$ Hz$, 16$ kHz$]$} are hardly audible by humans and they are not considered. Finer tuning of these bands is left for future works.

In each sub-band $B$, $kurt^{B}_{Nin}(\ell)$ and $kurt^{B}_{Nout}(\ell)$ are computed by substituting $X(k, \ell)$ in Eq.\,\ref{eq:inst_kurt} respectively by $N_{in}^{+}(k, \ell)$ and $N_{out}^{+}(k, \ell)$ with $k \in B$. Moreover, sub-band energy-weights $w^B(\ell)$ are computed as follows:
\begin{equation}
w^B(\ell) = 10 \log_{10} \left( \frac{1}{K_B} \sum_{k \in B} 10 \verb|^|\left( N_{out}^{+}(k, \ell)  /10 \right) \right),
\end{equation}
where $K_B$ is the number of frequency bins in sub-band $B$.
It is thus proposed to turn the attention to the sub-band absolute instantaneous log-kurtosis ratio, defined as:
\begin{equation}
\Delta kurt^{B} (\ell) =\left| \ln \left(  \frac{ kurt^{B}_{Nout}(\ell) }{ kurt^{B}_{Nin}(\ell) }  \right) \right|.
\end{equation}

It is worth noting that the previous works (Sec.\,\ref{subsec:blackbox}) use the time average to compute $kurt_{Nout}$ and $kurt_{Nin}$, while here $\Delta kurt^B(\ell)$ is still time-dependent, allowing for an energy-weighted temporal mean, as explained in the following.

\subsection{Perceptually Improved (PI) Log-Kurtosis Ratio}

The proposed final measure is thus computed as the energy-weighted mean of $\Delta kurt^{B} (\ell)$:
\begin{equation}
\Delta kurt^{PI} = \frac{ 1 }{ W }   \sum_{\ell=1}^{L} \left( w^{B=b}(\ell)  \Delta kurt^{B=b} (\ell)  \right),
\end{equation}
where $W = \sum_{\ell=1}^{L} w^{B=b}(\ell)$, and $B\myeq b$ indicates that one band is selected. The selected band is the one for which $\sum_{\ell=1}^{L} \left(  w^B(\ell) \Delta kurt^{B} (\ell) \right)$ is maximum. This is the band for which the changes in kurtosis are more evident, taking into consideration also the sub-band energy. The band selection aims to resemble the quality perception of a listener, which would be dominated by the worst sounding sub-band.
For the example in Fig.~\ref{fig:explain}, this sub-band is $B\myeq 2$, i.e., $(0.75, 6]$ kHz, see Fig.\ \ref{fig:explain}(d) and (e). Fig.~\ref{fig:explain}(f) depicts the resulting $\Delta kurt^{B} (\ell)$ and $\Delta kurt^{PI}$.

In this implementation, $\Delta kurt^{B} (\ell)$ is limited to $0.5$ and $\Delta kurt^{PI}$ is linearly rescaled to the range $[0, 100]$ in the following evaluation (but not in Fig.\,\ref{fig:explain}).

\section{Evaluation}  
\label{sec:evaluation}

This section evaluates the performance of the newly proposed $\Delta kurt^{PI}$ and compares it with the baseline kurtosis-based measures $\Delta kurt$ and $\Delta kurt^W$ as well as with the APS. 
The response to gradually introduced musical noise is studied in Sec.\,\ref{sub:controlled}, while a correlation analysis with perceptual scores is carried out in Sec.\,\ref{sub:subj}.

\subsection{Controlled Artificial Distortion}
\label{sub:controlled}

As a first performance evaluation without the need for perceptual data, musical noise is artificially introduced in a gradual and controlled manner. In order to do so, a controlled percentage of randomly selected bins of the spectrogram are set to zero.
The percentage ranges from $0$\% to $99.8$\%.
It is expected that the tested measures exhibit three properties with respect to the distortion control parameter, i.e., monotonicity, low inter-item deviation, and high range spanning. The response score $\rho$ measures to what extend these properties are met \cite{torcoli:2016}. 
In order to be able to calculate $\rho$, $\Delta kurt$ is limited between $0$ and $1.4$, and $\Delta kurt^W$ between $0$ and $2.2$, where the upper limits correspond to the highest scores obtained in this test. The outputs are then linearly rescaled to the range $[0, 100]$.
The test signals are 11 mixtures of stereo background signals and speech recordings (female and male speakers) panned to the center. Each signal is 10 seconds long. The backgrounds comprise music, environmental recordings, and direct signals.

Fig.\,\ref{fig:controlled} depicts the response of the measures to the introduced musical noise via the mean over all test signals (solid lines) plus/minus the standard deviation (dotted lines). It can be observed that $\Delta kurt^{PI}$ outperforms the other kurtosis-based measures, especially in terms of monotonicity and inter-item deviation. 

\begin{figure}[t]
  \centering
  \centerline{\includegraphics[width=1\columnwidth]{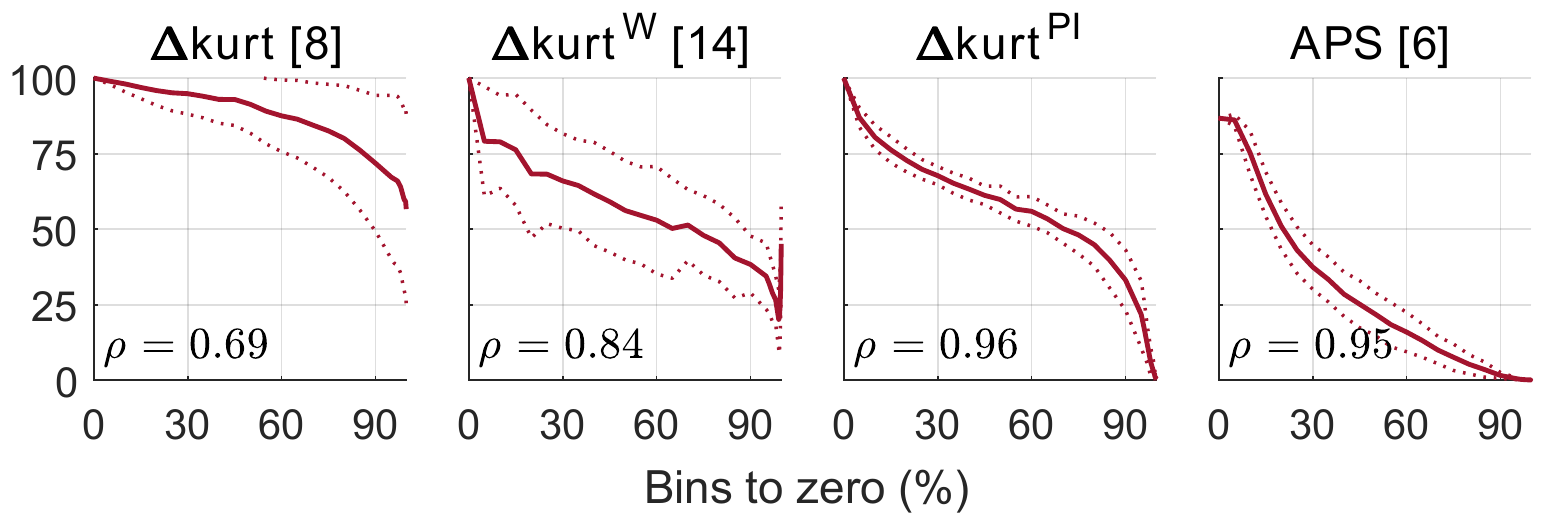}}
\caption{\label{fig:controlled}Responses to gradually introduced musical noise. The distortion control parameter is the percentage (from $0$ to $99.8$\%) of spectrum bins set to zero. The response score $\rho$ \cite{torcoli:2016} is also reported.}
\end{figure}

\subsection{Correlation Analysis with Perceptual Scores}
\label{sub:subj}

The perceptual scores from the 2 listening tests described in \cite{dick:2017} and \cite{emiya:2011} are here used as ground truth data to evaluate the correlation of the considered measures with the human perception of musical noise.
Figs.\,\ref{fig:birdies}-\ref{fig:peassDB} plot the average perceptual scores against the output of the measures. Pearson's correlation $r$ and  Kendall's rank correlation $t$ are also reported.
The following subsections introduce the perceptual scores and comment on the correlation of the considered measures with these scores.

\begin{figure*}[ht]
  \centering
  \centerline{\includegraphics[width=1.95\columnwidth]{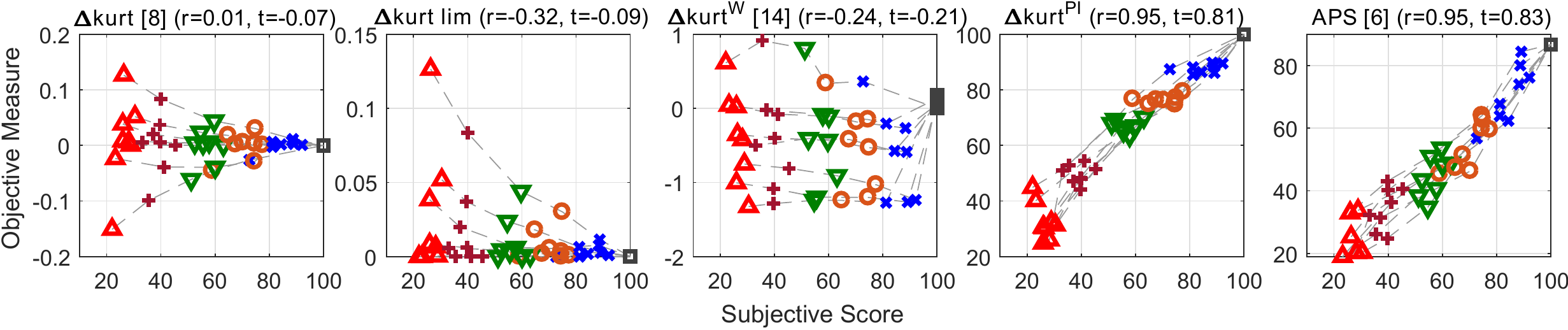}}
\caption{\label{fig:birdies}Correlation analysis with perceptual scores from the listening test on musical noise in audio coding\cite{dick:2017}. Different symbols are used for the different quality levels, while the dashed lines connect the data points corresponding to the same items in the different quality levels. 
The x-axis corresponds to average perceptual score, while the y-axis reports the output from the measure mentioned in the title of each plot, where Pearson's linear correlation coefficient $r$  and Kendall's rank correlation coefficient $t$ are also given. The reference signals (scoring 100 points) depicted with black squares are not considered in the calculation of $r$ and $t$, as they bias the coefficients towards higher values. }
\end{figure*}
\begin{figure*}
\centering
  \centerline{
\includegraphics[width=1.95\columnwidth]{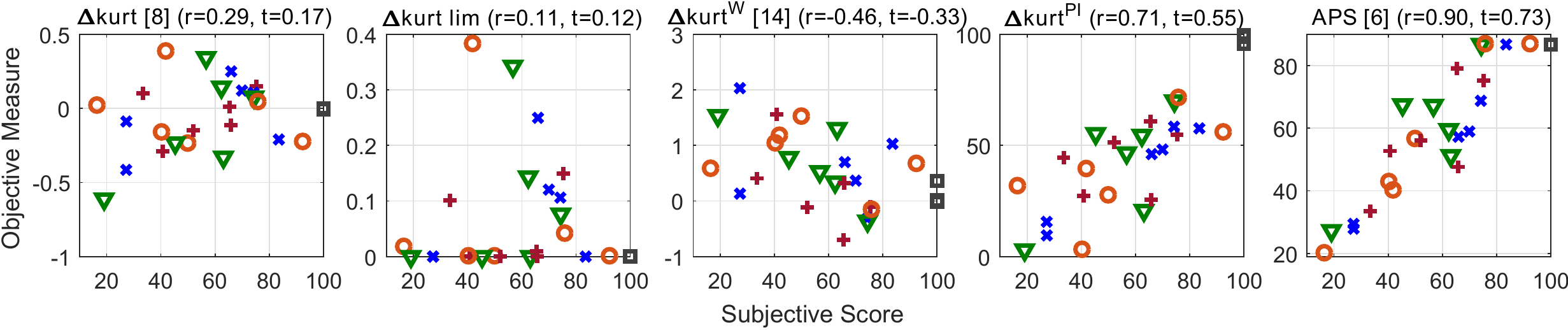}}
\caption{\label{fig:peassDB} Correlation analysis with perceptual scores from the listening test on musical noise in source separation, i.e., the training listening test for the APS~\cite{emiya:2011}. Different symbols are used for the different source separation algorithms. The x-axis corresponds to average perceptual score, while the y-axis reports the output from the measure mentioned in the title of each plot. The references are depicted with black squares, but they are not considered in the calculation of $r$ and $t$.}
\end{figure*}

\subsubsection{Listening Test on Audio Coding}
\label{subsub:audCod}

In \cite{dick:2017}, extensive subjective data was gathered; $16$ subjects assessed the quality of signals that were distorted in a controlled fashion with different monaural coding artifacts. 
The listening test was designed so that a process simulating sub-optimal audio coding distorted $8$ test items at $5$ different coarse quality levels. 
The perceptual scores are here used as perceptual reference scores, as also done in \cite{torcoli:2018comparing}, where the performance of 11 well-known objective evaluation tools was studied. The APS was found to be the measure with the highest correlation with the scores regarding musical noise. In~\cite{torcoli:2018comparing}, $\Delta kurt$ was limited to zero. This is herein referred to as $\Delta kurt$ $lim$.

Ideally, audio coding does not introduce any perceivable change. Hence, for this application, the entire input and output signals can be considered as $Nin$ and $Nout$, without the need for selecting the noise components only.

As it can be observed in Fig.\,\ref{fig:birdies}, $\Delta kurt^{PI}$ clearly correlates with the perceptual scores ($|r|=0.95$ and $|t|=0.81$), nearly as well as the APS. The APS is however computationally very expensive. 

The PEASS Matlab implementation \cite{emiya:2011} takes about 20 seconds for computing the APS of 5 seconds of audio on Matlab 2016 using a quad-core processor Intel i7-3770 at 3.40~GHz without any other competing process running. Under the same conditions, the author's non-optimized Matlab implementation of $\Delta kurt^{PI}$ requires about 0.1 seconds, i.e., it has two orders of magnitude lower computational complexity than APS.

Moreover, the proposed $\Delta kurt^{PI}$ clearly outperforms the baselines based on spectral kurtosis, i.e., $\Delta kurt$, $\Delta kurt$ $lim$, and $\Delta kurt^W$, which exhibit from low to no correlation with the perceptual scores. 
$\Delta kurt^{PI}$ outperforms also all the other state-of-the-art tools considered in \cite{torcoli:2018comparing} for this task.

\subsubsection{Listening Test on Source Separation}
\label{subsub:BSS}

The PEASS toolbox (and so the APS) was trained on publicly available results of listening tests focusing on different quality aspects~\cite{emiya:2011}. 
In one of these, 23 listeners (3 of them were excluded in a screening phase) rated the quality of 10 signals processed by 4 source separation algorithms in terms of \textit{absence of additional artificial noise}. 
The results from this test are here used as perceptual reference scores.
The content of the signals is very heterogeneous, e.g., multiple overlapping speakers or musical ensembles with or without singing voice. The items have different sampling frequencies, i.e., 16 or 44.1 kHz (resampling to 48 kHz is performed). 
The results exhibit significant noise, as 10 listeners of the considered 20 failed to detect the hidden reference.
For the APS calculation, the reference target and interfering sources are used (i.e., fully \textit{intrusive} approach). On the other hand, in order to calculate the kurtosis-based measures, simply the activity of the reference target source is used to select the time frames where only the interfering sources are present, i.e., $Nin$ and $Nout$. 
Items 1, 2, and 4 are not considered here, as the target source is always active. For the other items, the time frames available to the kurtosis-based measures are about the 15\% of the ones available to the APS. 
The anchor signals are not considered, as they were created adding musical noise to the target sources only. 
The noisiness of the perceptual scores, the limited frame number, and the fact that the APS was trained on these scores make the APS a particularly challenging benchmark for this dataset. 

As it can be observed in Fig.\,\ref{fig:peassDB}, $\Delta kurt^{PI}$ exhibits lower correlation for this test ($|r|=0.71$ and $|t|=0.55$), but it clearly outperforms the kurtosis-based baselines. 

\section{Conclusion}
This paper proposed a novel measure of musical noise based on spectral kurtosis and on perceptually motivated extensions. The measure was shown to clearly outperform the state-of-the-art kurtosis-based measures both in terms of response score and in terms of correlation with human perception. As perceptual reference scores, the results from two listening tests were considered.

When considering the listening test on which the APS was not trained, the proposed new measure performs nearly as well as the APS, which was found to be the best available measure for this task. Furthermore, the new measure requires two orders of magnitude lower computational complexity than APS.

\section{ACKNOWLEDGMENT}
Sincere thanks go to Jouni Paulus, Christian Uhle, J{\"u}rgen Herre, Sascha Dick, and Matthew Jefferson for the precious discussions.

\balance
\bibliographystyle{IEEEtran}
\bibliography{./Bib}

\end{sloppy}
\end{document}